\documentclass[traditabstract]{aa}
\usepackage{txfonts}
\usepackage{graphicx}
\usepackage{color}
\usepackage{placeins}

\begin{document}

\title{Galactic Pal-eontology: \\ abundance analysis of the disrupting globular cluster Palomar 5}

\author{
Andreas Koch\inst{1},   
\and 
Patrick C\^ot\'e\inst{2}
}
\authorrunning{A. Koch \& P. C\^ot\'e}
\titlerunning{Spectroscopy of Palomar 5}
\offprints{A. Koch; \email{a.koch1@lancaster.ac.uk;}}
\institute{
Department of Physics, Lancaster University, Lancaster LA1 4YB, United Kingdom
\and
National Research Council of Canada, Herzberg Astronomy and Astrophysics Program, 5071 West Saanich Road, Victoria, BC V9E 2E7, Canada
}
\date{}
\abstract{
We present a chemical abundance analysis of the tidally disrupted globular cluster (GC) Palomar 5. By co-adding high-resolution spectra of 15 member stars from the cluster's main body, taken at 
low signal-to-noise with the Keck/HIRES spectrograph, we were able to measure integrated 
abundance ratios of 24 species of 20 elements including all major nucleosynthetic channels
(namely the light element Na; $\alpha$-elements Mg, Si, Ca, Ti; Fe-peak and heavy elements Sc, V, Cr, Mn, Co, Ni, Cu, Zn; and the neutron-capture elements 
Y, Zr, Ba, La, Nd, Sm, Eu). 
The mean metallicity of $-1.56\pm0.02\pm0.06$ dex (statistical and systematic errors)
agrees well with the values from individual, low-resolution measurements of individual stars, but it is lower than previous high-resolution results of a small number of stars in the literature. Comparison with Galactic halo stars and other disrupted and unperturbed GCs
renders Pal~5 a typical representative of the Milky Way halo population, as has been noted before, 
emphasizing that the early chemical evolution of such clusters is decoupled from their later dynamical history. 
We also performed a test as to the detectability of light element variations in this co-added abundance analysis technique 
and found that this approach is not sensitive even in the presence of a broad range in sodium of $\sim$0.6 dex, a value typically found in the old halo GCs. Thus, while  methods of determining the global abundance patterns of such objects are well suited to study their overall enrichment histories, chemical distinctions of their multiple stellar populations is still best obtained from 
measurements of individual stars.
}
\maketitle
%
%
%
%
\section{Introduction}
Globular clusters (GCs) are amongst the oldest objects in the Universe and offer unique testbeds to probe 
internal evolutionary processes at those early times, as well as the build-up of the Galactic halo. 
As to the first point, it is now well-established that any GC observed to date hosts several stellar populations distinct in 
age, their light-element abundances { (e.g., Piotto et al. 2007; Gratton et al. 2012; Milone et al. 2017)}, and, in a few cases, 
in some heavy elements (e.g., Marino et al. 2015; Roederer et al. 2016). 
{ In the latter case, those GCs host a fraction of stars usually differing also in their iron content, and in each of the metallicity subgroups
there is evidence of variations in the light elements (e.g., Carretta et al. 2010; Marino et al. 2011). 
These light-element variations point} to proton-capture reactions acting in the hot interiors of stars associated with the first generations to have formed within the clusters. 
Even in observations with low-number statistics that do not allow for a representative sampling of the omnipresent Na-O anti-correlation (Carretta et al. 2009),
abundance anomalies can be found such as spreads in heavy elements,  
that { indicate  complex enrichment histories} 
(e.g., Koch \& McWilliam 2014; Hanke et al. 2017). 

Secondly,  comparisons of the global chemical element abundances of GCs with those of the stellar Galactic halo 
have implications for the accretion history of the Milky Way. In this regard, 
several candidates have been chemically and kinematically identified as having originated in the disrupted 
Sagittarius dwarf galaxy (e.g., Law \& Majewski 2010; Sbordone et al. 2015). 
Moreover, by studying light-element (in particular CN and CH) variations in GC and halo stars, 
the fraction of the halo that was donated by disrupted satellites can be efficiently constrained (e.g., 
Martell \& Grebel 2010).

Palomar 5 (hereafter Pal~5) is  one of the most prominent examples of a stellar system in severe tidal disruption, having lost 
up to 90\% of its mass into tidal tails that stretch over 10$\degr$ across
 the sky (Odenkirchen et al. 2002, 2003). 
Previous photometric (e.g., Sandage \& Hartwick 1977; Dotter et al. 2011) and  low-resolution spectroscopic 
studies  have identified it as a moderately metal-poor ([Fe/H]$\sim -1.4$ dex) system
(Smith 1985; Kuzma et al. 2015; Ishigaki et al. 2015). The latter two studies extended their analysis over broad parts of the tidal streams 
and they recovered a strong radial velocity gradient that is in line with simulations of the tidal disruption and can serve as input 
to improving halo mass models (Dehnen et al. 2004; Odenkirchen et al. 2009; Pearson et al. 2015).
However, no  metallicity gradient across the tails has been found. Smith et al. (2002) performed the first and only measurements of 
four stars at  { high} spectral resolution and high signal-to-noise ratio (SNR), suggesting that 
Pal~5 resembles the Galactic halo and higher-mass GCs in many regards; one exception was a lower-than-average
[$\alpha$/Fe] abundance ratio, which, at $\sim$0.16  dex, was found to lie significantly below the plateau value of $\sim$0.4 dex
inhabited by GC and field stars at similar metallicities. Furthermore, Smith (1985) and Smith et al. (2002) detected large 
abundance variations in the light elements (C, N, Na, and Al) and concluded 
that  whatever evolutionary processes are responsible for those variations{ in strongly disrupted clusters} are oblivious to the GC's present-day mass 
and that they were already imprinted at early times when the clusters still retained their initial mass.  
{ Conversely, Carretta et al. (2010) found that the 
extent of the variations in light elements due to proton-capture processing is tightly related to the 
present-day  total mass of clusters, which is the main driving parameter.} 

Many remote Milky Way GCs are faint systems, which leads to time-expensive observation strategies; 
thus one path to obtaining detailed chemical abundance information is to use integrated light spectroscopy (e.g., McWilliam \& Bernstein 2008)
or to co-add individual spectra at low SNR to emulate a higher-quality spectrum of a known underlying stellar population (Koch et al. 2009; Koch \& C\^ot\'e 2010).  
Following our work on the outer halo clusters Pal~3  and Pal~4, we now turn to the 
closer (R$_{\odot}$=23 kpc)  object Pal~5.

In Section 2 we describe our target list and observations, followed by a discussion of the radial velocity measurements and cluster membership assessments in Sect.~3.
Details on the co-added abundance determination and error analysis are given in Sect.~4 and we present the results in Sect.~5, with a focus on comparison with other GCs in Sect.~6. 
In Sect.~7 we briefly comment on the detectability of abundance variations with our approach, before concluding in Sect.~8. 
\section{Targets and observations}
%
The Pal~5 data discussed here were taken as part of a  broader program to study the internal dynamics of outer halo GCs 
(see, e.g., C\^ot\'e et al. 2002, Jordi et al. 2009, Baumgardt et~al. 2009; Frank et al. 2012). Our Pal~5 target stars were chosen from the 
red giant branch (RGB) and 
asymptotic giant branch  (AGB) sequences identified in the early photometric studies of Sandage \& Hartwick (1977; SH77) and the 
unpublished photometry and astrometry from Cudworth, Schweitzer, and Majewski (CSM; see Schweitzer et al. 1993).
Our target stars reach out to $\sim$2 half-light radii and all lie well within the cluster's tidal radius, avoiding the cluster's tidal features
(Odenkirchen et al. 2003).  Of course, the concept of tidal radius for such a highly disturbed object 
is largely meaningless and we refrain from investigating possible spatial trends in kinematics or 
abundance. Properties of the target stars and HIRES observations are given in Table~1. In Fig.~1, we show their location in a color-magnitude 
diagram (CMD) based on photometry from the Sloan Digital Sky Survey (SDSS; Alam et al. 2015). 
\begin{table}[htb]
\caption{Log of target stars.}
\centering
\begin{tabular}{ccccc}
\hline\hline       
 & $\alpha$  & $\delta$  & Exp. time  & S/N\tablefootmark{b} \\
Star\tablefootmark{a} & (J2000.0) &  (J2000.0) &  [s] & [pixel$^{-1}$] \\
\hline                  
CSM-003   & 15 16 15.93 & $-$00 09 28.79 & 180 & 6 \\  
CSM-029   & 15 16 13.99 & $-$00 09 33.39 & 300 & 7 \\ 
CSM-030   & 15 16 16.37 & $-$00 10 29.76 & 300 & 6 \\ 
CSM-032   & 15 16 09.52 & $-$00 02 39.64 & 240 & 7 \\ 
CSM-045   & 15 16 20.85 & $-$00 08 42.72 & 180 & 9 \\ 
CSM-174   & 15 16 07.67 & $-$00 10 18.64 & 180 & 7 \\ 
CSM-598   & 15 16 19.18 & $-$00 11 31.11 & 600 & 8 \\ 
SH77-E    & 15 15 58.83 & $-$00 05 17.02 & 180 & \llap{1}0 \\ %
SH77-F    & 15 15 56.05 & $-$00 06 05.67 & 180 & \llap{1}2 \\ %
SH77-G    & 15 16 08.61 & $-$00 08 03.19 & 180 & 8 \\ 
SH77-H    & 15 15 52.54 & $-$00 07 40.47 & 180 & 7 \\
SH77-J    & 15 15 49.64 & $-$00 07 00.73 & 240 & 7 \\
SH77-K    & 15 16 06.47 & $-$00 07 00.91 & 240 & 7 \\
SH77-L    & 15 16 01.95 & $-$00 08 02.75 & 240 & 8 \\
SH77-M    & 15 15 46.08 & $-$00 09 08.42 & 240 & 6 \\
SH77-N    & 15 15 59.45 & $-$00 08 59.69 & 240 & 6 \\
SH77-14   & 15 16 08.26 & $-$00 07 38.01 & 600 & 7 \\
SH77-U    & 15 15 54.73 & $-$00 06 54.95 & 300 & 5 \\
\hline                  
\end{tabular}
\tablefoot{
\tablefoottext{a}{Star-IDs from SH77 and CSM.}
\tablefoottext{b}{Given at 6600~\AA.}}
\end{table}
\begin{figure}[htb]
\begin{center}
\includegraphics[angle=0,width=1\hsize]{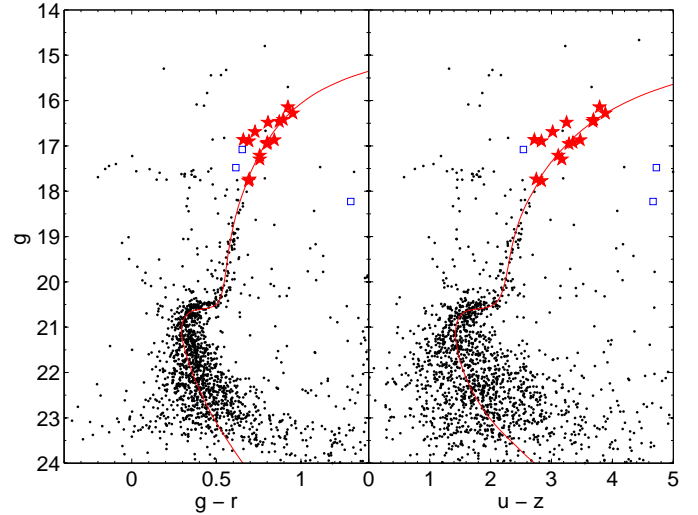}
\end{center}
\caption{CMD using SDSS photometry of stars within two half-light radii. Blue squares indicate foreground stars, open red symbols are AGB candidates, while our RGB targets are shown as solid stars. Dartmouth isochrones of the GC's parameters are shown as red lines.}
\end{figure}
\section{Radial velocities and membership}
Our observations were taken on May 31, 1998 using the HIRES echelle spectrograph (Vogt et al. 1994) on  the Keck I telescope with the C1 decker 
(slit width of 0.86$\arcsec$) and 1$\times$2 binning, which provides a spectral resolution of $R$=45000{ and a full spectra coverage of 
4300--6720 \AA.} 
The data were reduced with the Makee\footnote{MAKEE was developed by T. A. Barlow specifically for reduction of Keck HIRES data. It is freely available on the World Wide Web at the Keck Observatory home page, http://www2.keck.hawaii.edu/inst/hires/makeewww.} pipeline. 
The radial velocity of each star was measured by cross-correlating 
its spectrum against that of a master template created during each run (see below) from observations of the International Astronomical Union's (IAU)
standard stars. In order to
minimize possible systematic effects, a master template for each observing run was derived from an
identical subsample of IAU standard stars. From each cross-correlation function, we measured both $v_{HC}$, the heliocentric radial
velocity, and $R_{TD}$, the Tonry \& Davis (1979) estimator of the strength of the cross-correlation peak. 

During our broader Keck program, which spanned seven observing runs 
in 1998 and 1999 (13 nights in total), we obtained 53 distinct radial velocity measurements for 23 different RGB and subgiant stars belonging to 
Pal 3, Pal 4, Pal 5, Pal 14, NGC 7492, and NGC 2419. Using these repeat measurements and
following the procedures described in Vogt et al. (1995), we derived an empirical relationship between
our radial velocity uncertainties, $\epsilon(v_{HC})$, and the strength of cross-correlation peak: $\epsilon(v_{HC}) = \alpha / (1+R_{TD})$, 
where  $\alpha=9.0^{+2.4}_{-1.6}$ km/sec (90\% confidence limits). The final radial velocities and their uncertainties
are recorded in Table~2.  
\begin{table*}[htb]
\caption{Properties of the target stars.}
\centering
\begin{tabular}{cccccccccccc}
\hline\hline       
Star & $u$ & $g$ & $r$ & $i$ & $z$ & v$_{\rm HC}$ [km\,s$^{-1}$] &Type\tablefootmark{a}  &
T$_{\rm eff}$ [K] & log\,$g$ & $\xi$ [km\,s$^{-1}$] & [Fe/H]$_{\rm MgT}$\\
\hline                  
CSM-003   & 18.934 & 16.869 & 16.027 & 15.673 & 15.461 & $-52.19\pm$0.50  &  RGB & 4473 & 1.40 & 1.76 & $-1.32$ \\
CSM-029   & 19.205 & 17.296 & 16.538 & 16.217 & 16.033 & $-57.42\pm$0.54  &  RGB & 4607 & 1.67 & 1.68 & $-1.33$ \\
CSM-045   & 18.564 & 16.279 & 15.327 & 14.921 & 14.679 & $-59.28\pm$0.37  &  RGB & 4317 & 1.03 & 1.86 & $-1.45$   \\
SH77-F    & 18.622 & 16.419 & 15.524 & 15.157 & 14.933 & $-58.51\pm$0.40  &  RGB & 4410 & 1.16 & 1.80 & $-1.43$   \\
SH77-G    & 18.681 & 16.464 & 15.592 & 15.217 & 15.000 & $-58.28\pm$0.45  &  RGB & 4418 & 1.19 & 1.80 & $-1.42$  \\
SH77-K    & 18.950 & 16.926 & 16.125 & 15.783 & 15.597 & $-59.25\pm$0.46  &  RGB & 4531 & 1.46 & 1.73 & $-1.33$ \\
SH77-L    & 18.905 & 16.955 & 16.155 & 15.816 & 15.616 & $-58.90\pm$0.47  &  RGB & 4536 & 1.48 & 1.72 & $-1.38$ \\
SH77-N    & 19.057 & 17.210 & 16.453 & 16.128 & 15.943 & $-60.32\pm$0.59  &  RGB & 4601 & 1.63 & 1.68 & $-1.33$  \\
SH77-14   & 19.437 & 17.772 & 17.082 & 16.787 & 16.601 & $-58.41\pm$0.53  &  RGB & 4731 & 1.94 & 1.60 & $-1.32$ \\
SH77-U    & 19.329 & 17.735 & 17.040 & 16.744 & 16.578 & $-59.09\pm$0.90  &  RGB & 4722 & 1.92 & 1.61 & $-1.13$ \\
\hline
CSM-032   & 18.585 & 16.894 & 16.201 & 15.911 & 15.749 & $-59.00\pm$0.58  &  AGB & 4737 & 1.47 & 1.60 & $-1.44$  \\
CSM-174   & 18.410 & 16.481 & 15.676 & 15.344 & 15.161 & $-61.14\pm$0.50  &  AGB & 4545 & 1.17 & 1.72 & $-1.49$ \\
SH77-E    & 18.393 & 16.142 & 15.219 & 14.830 & 14.601 & $-57.48\pm$0.38  &  AGB & 4358 & 0.88 & 1.83 & $-1.46$ \\
SH77-H    & 18.514 & 16.690 & 15.961 & 15.648 & 15.493 & $-57.24\pm$0.58  &  AGB & 4653 & 1.33 & 1.65 & $-1.37$ \\
SH77-J    & 18.466 & 16.863 & 16.203 & 15.911 & 15.744 & $-57.38\pm$0.62  &  AGB & 4773 & 1.48 & 1.57 & $-1.58$ \\
\hline
CSM-030   & 21.280 & 17.481 & 16.865 & 16.533 & 16.558 & $-38.32\pm$0.57  &  FG & \ldots & \ldots & \ldots & \ldots \\
CSM-598   & 20.772 & 18.226 & 16.932 & 16.409 & 16.104 & $-14.33\pm$0.61  &  FG  & \ldots & \ldots & \ldots & \ldots  \\
SH77-M    & 18.500 & 17.079 & 16.426 & 16.131 & 15.959 & \phantom{$-$}$20.18\pm$1.11  &  FG   & \ldots & \ldots & \ldots & \ldots \\
\hline                  
\end{tabular}
\tablefoot{
\tablefoottext{a}{Stellar type based on CMD and spectral properties: red giant branch (RGB), asymptotic giant branch (AGB), or
foreground (FG) dwarf.}
}
\end{table*}

Several spectroscopic studies have since been carried out for Pal~5. 
All of the  four stars analyzed by  Smith et al. (2002), using Keck/HIRES at high SNR 
but{ lower resolution (R=34000\footnote{ This is the same value as used in our previous study of Pal~4; Koch \& C\^ot\'e (2010).}) than in the present work}, 
are included in our data set as well, whereas 
none of the stars are included in  Odenkirchen's et al. (2009) kinematic study.          
Odenkirchen et al. (2002) performed a kinematic study on their spectra obtained with the 
Ultraviolet and Visual Echelle Spectrograph (UVES) 
at high-resolution, but low SNR ($\sim$10 pixel$^{-1}$), 
of which ten stars are also in our present sample. 
We observed  11 stars  in common with 
the work of Kuzma et al. (2015), who obtained 
velocity and calcium-triplet metallicity measurements based on  low-resolution ($R\sim$10000) spectroscopy with the AAOmega multi-fibre instrument.
Finally, one of our stars is also included in the data of Ishigaki et al. (2016) who used low-resolution ($R\sim$7000) 
spectra to gather kinematics, metallicities, and [$\alpha$/Fe] ratios.

The mean heliocentric velocity of $-58.3\pm0.5$ km\,s$^{-1}$ we find for Pal 5 is in excellent agreement with all previous studies of the 
central regions to within the uncertainties. 
The dispersion of 1.8$\pm$0.3 km\,s$^{-1}$ from our data is slightly higher than
the literature values of typically 1.1$\pm$0.3 km\,s$^{-1}$, but this may be due to the lower number of stars we observed or to our confinement 
to the more central parts of the cluster. 
Three of our targets turned out to be non-members based on their strongly deviating velocities. 
While the presence of a strong velocity gradient throughout this tidal  system (e.g., Odenkirchen et al. 2002) could in principle render 
these stars related to Pal~5, the appearance of their spectral features such as broad, gravity-sensitive Mg triplet lines or the Ca 6162\AA-line, 
clearly indicates that they are foreground dwarfs.
As the CMD in Fig.~1 implies, five of the targets  are located on the AGB, as was also suggested for four of them by Odenkirchen et al. (2002). The separation of our sample into RGB, AGB, and foreground stars is indicated in Table~2. 
\section{Analysis}
The original aim in acquiring this data set was to study the kinematics of Galactic halo clusters (see Sect.~2)
and so the exposure times were chosen to be in the range of three--ten minutes, thus operating at low SNR (Table~1; see also Odenkirchen et al. 2002). While this SNR is adequate for measuring accurate radial velocities, a detailed chemical analysis of individual stars is precluded. Thus, we resort to our method of co-adding spectra in the manner outlined in Koch et al. (2009) and Koch \& C\^ot\'e (2010). 
\subsection{Stellar parameters}
Stellar effective temperatures of the individual stars were obtained from their (V$-$I) colors using the calibrations of Ram\'\i rez \& Mel\'endez (2005). To this end, we transformed the  SDSS $g,r,i$ magnitudes to the Johnson-Cousins system following the prescriptions of Jordi et al. 
(2006) and adopting a cluster reddening of E(B$-$V)=0.08 mag (Dotter et al. 2011).{ Although this is larger than the value listed by Harris (1996 [2010 edition]), 
it is more in line with the reddening maps of Schlegel et al. (1998) and Schlafly et al. (2011), who list 0.05 and 0.06 mag, respectively,  and it also gave the best representation of the SDSS CMD in Fig.~1.}
The typical uncertainty on T$_{\rm eff}$ due to the 
(small) photometric errors in the SDSS and the calibrations is $\sim$50 K.

Next, the surface gravities were obtained using the standard equations of stellar structure, where we used the above temperature, 
a stellar mass of  0.8 M$_{\odot}$ for the RGB and 0.6 M$_{\odot}$ for AGB stars, and a distance to Pal 5's center of 23.2 kpc. 
As an initial estimate of the stars' metallicity that enters the bolometric corrections, we adopted the photometric GC mean of $-1.35$ dex. 
One uncertain factor is the pronounced presence of mass segregation within Pal 5 and throughout its tails 
(Koch et al. 2004), which could also affect the analysis of our stars that are located throughout the cluster.
Propagating the errors on all the above quantities translates into typical gravity uncertainties of 0.15 dex. 

Similar to our previous work (Koch et al. 2009; Koch \& C\^ot\'e 2010), 
we derived microturbulence velocities, $\xi$, using an empirical calibration of $\xi$ with T$_{\rm eff}$ based on the halo stars of 
Roederer et al. (2014). 
{ This relation reads: ${4.567 - 6.2694\,\times\,10^{-4}\, \times {\rm  T_{eff}}}$. 
}
The inferred uncertainty from the scatter around this relation is 0.10 km\,s$^{-1}$. 

While a global cluster metallicity is available for Pal~5 (e.g., Dotter et al. 2011) and previous low-resolution studies have
measured metallicities of individual stars (e.g., Kuzma et al. 2015; see also \S5.1), we opted to 
obtain initial metallicities of our targets that enter the stellar atmospheres from 
the Mg\,{\sc i} line index. This index uses the strong Mg triplet lines at 
 5167 and 5173\AA~and is defined and calibrated on the scale of Carretta \& Gratton (1997) as in Walker et al. (2007) and Eq. 2 in 
 Koch et al. (2009).  For this, we assume a horizontal branch magnitude of V$_{\rm HB}$=17.51 mag (Harris 1996 [2010 edition]).
 { These values should only be taken as a general estimate for the metallicity distribution, since the Mg triplet 
 has also dependencies on temperature and gravity. In particular, we note a trend of [Fe/H]$_{\rm Mg}$ with effective temperature. 
 To identify possible causes, we consulted the Schlegel et al. (1998) maps  and SDSS photometry for signs of differential 
 reddening (following the approach of Milone et al. 2012; Kacharov et al. 2014). Either method indicates a very homogeneous reddening across the face of Pal~5 with a 1$\sigma$-scatter well below 0.01 mag. Similarly, the actual value of E(B$-$V) only leads to a systematic
 shift of our T$_{\rm eff}$-scale, which has no bearing on the purported trend with metallicity. 
 When accounting for  uncertainties in our measurements, the  correlation coefficient of metallicity with temperature 
 becomes 0.25$\pm$0.35 and we conclude that any possible trend is driven by statistical uncertainties. 
}
 
 All stellar parameters we have used in creating our stellar atmospheres are listed in Table~2 and plotted in Fig.~2. 
\begin{figure}[!tb]
\begin{center}
\includegraphics[angle=0,width=1\hsize]{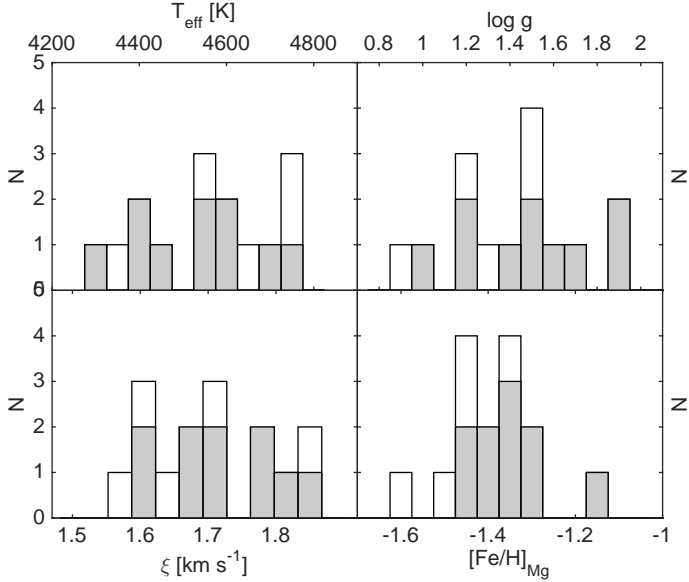}
\end{center}
\caption{Histogram of stellar parameters. Solid bars are for RGB stars and clear ones for the AGB subsample.}
\end{figure}
\subsection{Co-added abundance measurements}
Consistent with our previous work, we have median-combined the spectra, irrespective of evolutionary status, after weighting by their SNR.
{ This led to an SNR of $\sim$30 per pixel at 6600 \AA.} 
Equivalent widths (EWs) were measured from the line lists 
of Koch et al. (2016) and Ruchti et al. (2016) by fitting Gaussian profiles with the {\em splot} task within the Image Reduction and Analysis Facility (IRAF). These line lists and the EW measurements are presented in Table~3.
\begin{table}[htb]
\caption{Line list.}
\centering
\begin{tabular}{ccccccc}
\hline\hline       
$\lambda$ &  & E.P. &  &  \multicolumn{3}{c}{$\left<EW\right>$ [m\AA]} \\
\cline{5-7}
[\AA] & \raisebox{1.5ex}[-1.5ex]{Species} & [eV] &\raisebox{1.5ex}[-1.5ex]{log\,$gf$} & All & RGB & AGB \\
\hline                  
5682.633 & Na\,{\sc i} &   2.102  & $-$0.700  & 61.3 & 68.9 & 35.9 \\
5688.205 & Na\,{\sc i} &   2.104  & $-$0.404  & 91.9 & 96.0 & 38.4 \\
6154.225 & Na\,{\sc i} &   2.102  & $-$1.547  & 11.8 & \ldots  &10.2  \\
6160.747 & Na\,{\sc i} &   2.104  & $-$1.246  & 31.0 & 35.8  & \ldots \\
4571.096 & Mg\,{\sc i} &   0.000 & $-$5.623   & 149.6 & 132.1  & 162.1 \\
4702.991 & Mg\,{\sc i} &   4.346 & $-$0.440   & 171.5 & 162.7 & 157.2  \\
5711.088 & Mg\,{\sc i} &   4.346 & $-$1.724   & 76.62 & 61.2 & 77.6 \\
\hline                  
\end{tabular}
\tablefoot{Table~3 is available in its entirety in electronic form via the  Centre de Donn\'ees astronomiques de Strasbourg (CDS).}
\end{table}

In the following analysis, we have used the 2014 version of the stellar abundance code MOOG (Sneden 1973). 
From the stellar parameters derived in Sect.~4, we created individual stellar atmospheres for each star. Specifically, 
the ATLAS grid of Kurucz' one-dimensional 72-layer, plane-parallel, line-blanketed models without convective overshoot was interpolated 
assuming local thermodynamic equilibrium (LTE) for all species, together with the $\alpha$-enhanced opacity distribution functions, AODFNEW.
We then computed theoretical EWs for all transitions using MOOG's {\em ewfind} driver and combined them into a mean $\left<EW\right>$ 
by applying the same SNR-based weights as for the observed spectra (see, e.g., Eq.~1. in Koch \& C\^ot\'e 2010). 
Finally, the abundance ratio of each element was varied until the co-added $\left<EW\right>$ matched the observed EW for each line to yield an integrated abundance ratio.
\subsection{Abundance errors}
We quantified the measurement error by standard procedures: firstly, the statistical error 
is given by the 1$\sigma$ line-to-line scatter and the number of lines, N,  used to measure an element's abundance, 
each of which are listed in Table~4. For a few elements, only one weak feature was measurable and we varied the observed 
EW by 5 m\AA~ for those cases (which yielded reasonable fits to the line profile in {\em splot}).
This led to typical uncertainties in the respective abundances of 0.15 dex.
 
Secondly, to determine the systematic uncertainties, we varied
each of the stellar parameters by their typical uncertainty (T$_{\rm eff}\pm50$ K; $\log\,g\pm0.15$ dex; $\xi\pm0.10$ km\,s$^{-1}$) 
and re-ran the co-addition scheme. 
We note that this presupposes that all of the stars are affected by the same amount of error and the same sense of departure from the 
un-altered values. The difference for each element upon this variation in each parameter is listed in Table~4; there, we also indicate
the effect of switching from the $\alpha$-enhanced opacity distribution functions to the solar-scaled distributions, ODFNEW. 
Due to the strong correlations between temperature, gravity, and, as per our construction, the microturbulence, 
also the systematic uncertainties are not independent of each other. Thus, we caution that the total error we list in Table~4, which is 
merely the quadratic sum of all contributions, should be taken as  a conservative upper limit at most. 

In Koch \& C\^ot\'e (2010), we investigated  further error sources such as radial velocity uncertainties which can lead to additional 
line broadening, but found that this accounts for less than 0.04 dex in the error budget. Likewise, the 
accidental inclusion of foreground stars would add no more than 0.02 dex of abundance uncertainty. 
\section{Abundance results}
The results from our co-added abundance measurements are presented in { Table~4}. All our abundances are on the scale of solar abundances 
from Asplund et al. (2009). 
Figures~3--5 show these abundance ratios in comparison with other Galactic components, such as halo field stars (from Roederer et al. 2014), the bulge (Johnson et al. 2012, 2014), and the disks (Koch \& Edvardsson 2002; Bensby et al. 2014). 
Furthermore we added GC data for M 5 as a representative of an undisturbed system at comparable metallicity (Ivans et al. 2001), 
the outer halo GCs Pal~3 and Pal~4 (Koch et al. 2009; Koch \& C\^ot\'e 2010), 
and several objects that have been associated with the disrupting Sgr dwarf galaxy (see the caption of Fig.~3 for the color coding): 
Pal~12 (Cohen 2004); 
Terzan~7 (Sbordone et al. 2005); 
Arp~2 (Mottini et al. 2008); 
 Terzan~8 and the massive M~54 (Carretta et al. 2014), and 
NGC 5053 and 5634 (Sbordone et al 2015). 
Finally, the tidally disturbed cluster NGC~5466 (Lamb et al. 2015) is shown as an orange hexagon. 
%
%
%
%
\begin{table*}[htb]
\caption{Abundance results from co-added spectra, { where we also list results separately for the  AGB- and RGB-only subsamples.} Abundance ratios for ionized species are given relative to Fe\,{\sc ii}. 
For iron itself, [Fe/H] is listed. The line-to-line scatter $\sigma$ and number of measured lines, N, determine the statistical 
error, while the systematic uncertainties are indicated in the last five columns.}
\centering
\begin{tabular}{cccccccccccccccccc}
\hline\hline       
 & [X/Fe] & $\sigma$ & N  & & [X/Fe] & $\sigma$ & N  & & [X/Fe] & $\sigma$ & N  &
$\Delta$T$_{\rm eff}$ & $\Delta$log\,$g$ & $\Delta\xi$ &  &    \\
\cline{2-4}\cline{6-8}\cline{10-12}
\raisebox{1.5ex}[-1.5ex]{Species} & \multicolumn{3}{c}{All} & & \multicolumn{3}{c}{RGB} & & \multicolumn{3}{c}{AGB} &  
$\pm$$50$ K & $\pm$$0.15$ dex & $\pm$$0.1$ km\,s$^{-1}$ & \raisebox{1.5ex}[-1.5ex]{ODF} & \raisebox{1.5ex}[-1.5ex]{$\sigma_{\rm sys}$}  \\
\hline                   
Fe\,{\sc i}  & \llap{$-$}1.56 &  0.20 & 104 &  & \llap{$-$}1.65 & 0.26 & 95 & & \llap{$-$}1.66 & 0.29 & 83 & $\pm$0.05& $<0.01$ & $\mp$0.03 & $<0.01$  & 0.06 \\
Fe\,{\sc ii} & \llap{$-$}1.46 &  0.30 & 13  &  & \llap{$-$}1.50 & 0.45 & 12 & & \llap{$-$}1.64 & 0.60 & 13 & $\mp$0.06 & $\pm0.10$ & $\mp0.02$ & $-0.05$ & 0.13 \\
Na\,{\sc i}  &           0.44 &  0.16 &  4  &  &  0.65 &   0.08 &  3 & & 0.13 & 0.20 & 3 & $\mp$0.04 &  $\pm$0.01 &  $\pm$0.01 &  $-$0.01 &  0.04  \\
Mg\,{\sc i}  &           0.44 &  0.31 &  3  &  &  0.25 &   0.44 &  3 & & 0.67 & 0.15 & 3 & $\mp$0.04 &  $\pm$0.01 &  $\pm$0.02 &  $-$0.02 &  0.05  \\
Si\,{\sc i}  &           0.53 &  0.25 &  9  &  &  0.47 &   0.58 &  8 & & 0.54 & 0.27 & 9 &   $<$0.01 &  $\mp$0.02 &   $<$0.01 &  $-$0.01 &  0.02  \\
Ca\,{\sc i}  &           0.38 &  0.25 & 21  &  &  0.40 &   0.26 & 21 & & 0.36 & 0.29 & 21 & $\mp$0.06 &  $\pm$0.01 &  $\pm$0.05 &  $-$0.04 &  0.09  \\
Sc\,{\sc i}  &           0.25 &  0.07 &  2  &  &  0.38 &   0.23 &  2 & & \ldots & \ldots & 0 & $\mp$0.09 &  $\pm$0.01 &   $<$0.01 &    $<$0.01 &  0.09  \\
Sc\,{\sc ii} &           0.09 &  0.13 &  6  &  &  0.16 &   0.10 &  6 & & 0.28 & 0.31 & 6 & $\pm$0.01 &  $\mp$0.06 &  $\pm$0.02 &  $-$0.02 &  0.07  \\
Ti\,{\sc i}  &           0.20 &  0.22 & 22  &  &  0.28 &   0.25 & 22 & & 0.47 & 0.38 & 18 & $\mp$0.09 &  $\pm$0.01 &  $\pm$0.03 &  $-$0.03 &  0.10  \\
Ti\,{\sc ii} &           0.22 &  0.28 & 10  &  &  0.28 &   0.41 & 10 & & 0.44 & 0.42 & 10 & $\pm$0.01 &  $\mp$0.06 &  $\pm$0.06 &  $-$0.06 &  0.10  \\
V\,{\sc i}   & \llap{$-$}0.03 &  0.21 &  9  &  &  0.01 &   0.21 &  8 & & 0.27 & 0.65 & 5 & $\mp$0.09 &  $\pm$0.01 &  $\pm$0.01 &  $-$0.01 &  0.09  \\
Cr\,{\sc i}  &          0.00 &  0.33 &  5  &  &  0.10 &   0.41 &  4 & & 0.12 & 0.49 & 5 & $\mp$0.09 &  $\pm$0.01 &  $\pm$0.03 &  $-$0.03 &  0.10  \\
Cr\,{\sc ii} &           0.10 & \ldots &  1  &  &  0.07 &  \ldots &  1 & & \ldots & \ldots & 0 & $\pm$0.03 &  $\mp$0.06 &  $\pm$0.02 &  $-$0.02 &  0.07  \\
Mn\,{\sc i}  & \llap{$-$}0.10 &  0.24 &  8  &  &  0.07 &   0.28 &  8 & & \llap{$-$}0.13 & 0.34 & 8 & $\mp$0.06 &  $\pm$0.01 &  $\pm$0.02 &  $-$0.02 &  0.06  \\
Co\,{\sc i}  & \llap{$-$}0.10 &  0.07 &  3  &  &  \llap{$-$}0.01 &   0.08 &  3 & & 0.18 & 0.08 & 2 & $\mp$0.05 &    $<$0.01 &  $\pm$0.01 &  $-$0.01 &  0.05  \\
Ni\,{\sc i}  &           0.05 &  0.24 & 18  &  &  0.12 &   0.17 & 17 & & 0.11  & 0.46 & 15 & $\mp$0.03 &  $\mp$0.01 &  $\pm$0.01 &  $-$0.02 &  0.04  \\
Cu\,{\sc i}  & \llap{$-$}0.49 &  0.22 &  2  &  &  \llap{$-$}0.21 &   0.01 &  2 & & \llap{$-$}0.18 & 0.15 & 2 & $\mp$0.06 &  $\mp$0.01 &  $\pm$0.03 &  $-$0.03 &  0.07  \\
Zn\,{\sc i}  &           0.05 &  0.05 &  2  &  &  0.04 &   0.38 &  2 & & \llap{$-$}0.10 & \ldots & 1 & $\pm$0.02 &  $\mp$0.04 &  $\pm$0.03 &  $-$0.03 &  0.06  \\
Y\,{\sc ii}  & \llap{$-$}0.35 &  0.05 &  2  &  &  \llap{$-$}0.32 &   0.03 &  2 & &  \llap{$-$}0.35 & 0.02 & 2 & $\mp$0.02 &  $\mp$0.05 &  $\pm$0.06 &  $-$0.07 &  0.11  \\ 
Zr\,{\sc ii} & \llap{$-$}0.13 &  0.10 &  2  &  & \ldots &  \ldots &  0 & & \ldots  & \ldots & \ldots &   $<$0.01 &  $\mp$0.06 &  $\pm$0.01 &  $-$0.02 &  0.06  \\
Ba\,{\sc ii} &           0.40 &  0.16 &  5  &  &  0.36 &   0.22 &  5 & & 0.55 & 0.26 & 5 & $\mp$0.02 &  $\mp$0.04 &  $\pm$0.06 &  $-$0.09 &  0.12  \\
La\,{\sc ii} &           0.13 & \ldots &  1  &  &  0.29 &  \ldots &  1 & & 0.26 & \ldots & 1 & $\mp$0.01 &  $\mp$0.06 &  $\pm$0.01 &  $-$0.01 &  0.06  \\
Nd\,{\sc ii} &           0.39 & \ldots &  1  &  &  0.32 &  \ldots &  1 & & \ldots & \ldots & \ldots & $\mp$0.01 &  $\mp$0.06 &  $\pm$0.02 &  $-$0.02 &  0.07  \\
Sm\,{\sc ii} &           0.24 &  0.10 &  4  &  &  0.41 &   0.22 &  4 & & 0.32 & 0.09 & 2 & $\mp$0.01 &  $\mp$0.06 &  $\pm$0.01 &  $-$0.01 &  0.06  \\
Eu\,{\sc ii} &           0.55 & \ldots &  1  &  & \ldots &  \ldots &  0 & & 0.51 & \ldots & 1 & $<$0.01 &  $\mp$0.06 &      $<$0.01 &  $-$0.01 &  0.06 \\
\hline                  
\end{tabular}
\end{table*}
\subsection{Iron abundance and metallicity}
The ensuing metallicity distribution from the Mg index (Fig.~2, bottom right) indicates a mean metallicity of $-1.38$ dex. This is in excellent agreement with both photometric values of $-1.4$ dex (Dotter et al. 2011) and, to within their (large) uncertainties, the low-resolution mean values of Kuzma et al. (2015) and Ishigaki et al. (2016) of $-1.48\pm0.10$ and $-1.35\pm0.06$, respectively. All values can be reconciled if 
we split our sample into RGB stars only ($-1.33\pm0.09$ dex) versus an AGB-only subset ($-1.47\pm0.08$ dex). 
{ We note, however, that the nominal uncertainties on this index measurement are typically large so that 
conclusions drawn from the distribution function should be taken with caution compared to the more reliable 
iron abundance from individual lines.} 
Smith et al. (2002) found a more metal-rich mean value, $-1.28\pm0.03$ dex, from four high-SNR, medium resolution spectra, { 
which is still consistent with the average [Fe/H] from our sample when only considering RGB stars}.  

From the co-added spectrum, we measured 104 Fe\,{\sc i} and 13 Fe\,{\sc ii} lines to arrive at an Fe-abundance that { is marginally lower than}
the aforementioned values, namely [Fe/H]=$-1.56\pm0.02\pm0.06$ dex (statistical and systematic errors, respectively) based on the neutral species. Ionization equilibrium is  marginally reached in this 
co-added approach, at [Fe\,{\sc i}/Fe\,{\sc ii}]=$-0.10\pm0.09$ dex (Table~4). In the following discussions, abundance ratios of ionized species will be referenced to the iron abundance from Fe\,{\sc ii} lines.
%
%
%
\subsection{Light elements: Na}
Unfortunately, both of the strong Na D lines fall on the gap between adjacent orders so we derived the Na-abundance from the weaker 
5682, 5688, 6154, and 6160 \AA~lines. No oxygen lines were detectable in the co-added spectrum.
At 0.44 dex, the mean, co-added [Na/Fe] abundance ratio of Pal~5 is high compared to halo stars at the same metallicity, which usually show solar to subsolar values. The corrections for Non-Local Thermodynamic Equilibrium (NLTE) in individual stars with
 stellar parameters as in our Pal~5 sample are on the order of $-$0.10 dex (Lind et al. 2011).
Generally, the presence of strong Na lines could provide evidence of Na-strong, second generation GC stars amongst the sample. 
In fact,  a pronounced light-element spread in Pal~5 of 0.3 dex (with a full range of 0.6 dex) was already noted by Smith et al. (2002). 
We further discuss the sensitivity of our analysis method to these light element variations in Sect.~6. 
\subsection{Alpha-elements: Mg, Si, Ca, Ti}
\begin{figure}[tb]
\begin{center}
\includegraphics[angle=0,width=0.8\hsize]{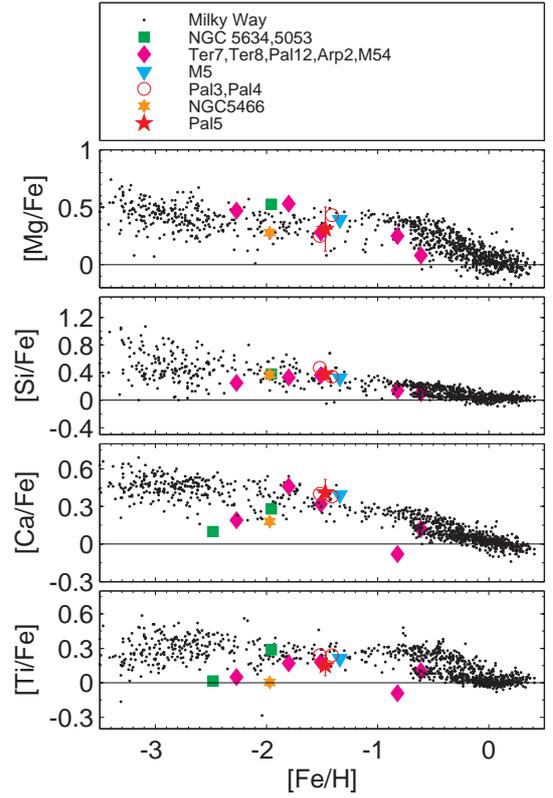}
\end{center}
\caption{Abundance results for the $\alpha$-elements. Literature data for the Milky Way (black dots) are from:  
halo -- Roederer et al. (2014); bulge -- Johnson et al. (2012,2014); disks -- Bensby et al. (2014).
Pal~5 is shown as a red star symbol, while other GCs are from Sbordone et al. (2015) -- NGC 5053, 5634 (green squares); 
Cohen (2004); Sbordone et al. (2007); Mottini et al. (2008);  Carretta et al. (2014) for 
Pal~12, Ter~7, Arp~2, Ter~8, and M~54 (magenta diamonds). Furthermore, M 5 is shown as a cyan triangle (Ivans et al. 2001), the outer halo GCs Pal~3 and Pal~4 as open circles (Koch et al. 2009; Koch \& C\^ot\'e 2010), and the disrupted NGC~5466 as an orange hexagon (Lamb et al. 2015).}
\end{figure}
Pal~5 shows an enhancement in the $\alpha$-elements, which overlaps with the halo distribution by merit 
of the latter's broad scatter.  A straight average over all four elements, Mg, Si, Ca, and Ti, yields [$\alpha$/Fe]=0.39 dex, 
although differences in each element's production channel renders other weighting schemes more appropriate.
Smith et al. (2002) found the abundances of these elements to be significantly 
 lower ($<$[Si,Ca,Ti/Fe]$>$=0.16, while we find 0.37 from those three elements) when compared to halo field stars 
or the similarly metal-poor GC M~5 (Ivans et al. 2001; Koch \& McWilliam 2010). 
Our data indicate a{  depletion from the $\alpha$-plateau by $\sim$0.2 dex, while the remainder of the elements, in particular Mg, 
lie very close the canonical value of $\sim$0.4 dex.
This difference is curious --  both Mg and Ti are $\alpha$-elements, but the Mg-production is 
also affected by proton-capture reactions, leading to
 the conversion into Al, while Ti is not affected by this channel.
 The lack of a significant Mg-depletion could then indicate that proton-capture processing in the first generation polluters for this cluster was
 rather inefficient.
Unfortunately, no Al could be measured owing to an SNR of $\sim$28 in the region of the commonly used 6696, 6698 \AA-lines.}
We note that both our measurement for Mg and Ti still overlap with the M~5-abundances (cyan triangles in Figs.~4--6) to within the errors. 
Our values are furthermore fully consistent with the measurements 
of individual stars in the  M~54, which,  at $-1.5$ has a very similar metallicity to Pal~5. 
This  massive GC  is thought to be the central cluster of the Sgr dwarf galaxy.
Finally, we note that there is an excellent  ionization balance between Ti\,{\sc i} and {\sc ii} when taken relative to Fe\,{\sc i} and Fe\,{\sc ii}. 
%
%
\subsection{Fe-peak and heavy elements: Sc, V, Cr, Mn, Co, Ni, Cu, Zn}
\begin{figure}[htb]
\begin{center}
\includegraphics[angle=0,width=0.8\hsize]{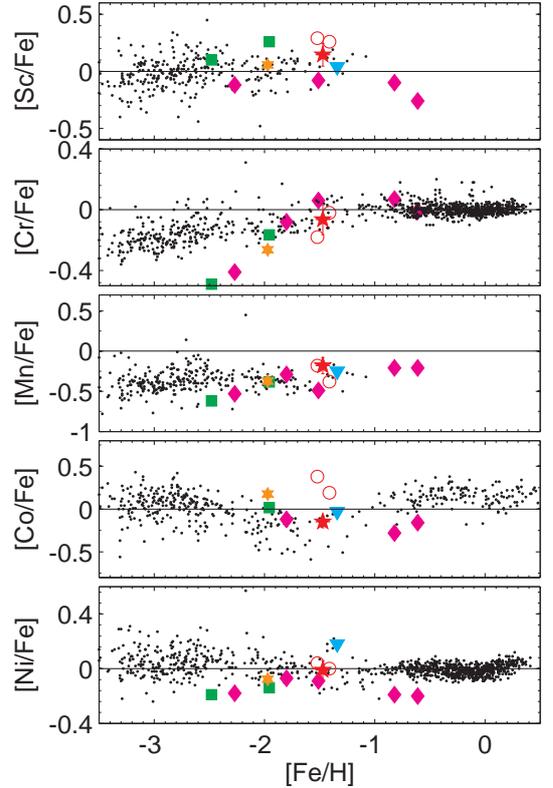}
\end{center}
\caption{Same as Fig.~3, but for Fe-peak and heavy elements.}
\end{figure}
None of the Fe-peak elements show a surprise as they all fall in the regime of the halo field star and GC 
distributions (Fig.~4). We note that ionization equilibrium is not fulfilled for Sc and Cr, but the measurements of their minority species 
are usually based on only one or two lines. 
\subsection{Neutron-capture elements: Zr, Y, Ba, La, Nd, Sm, Eu}
\begin{figure}[htb]
\begin{center}
\includegraphics[angle=0,width=0.8\hsize]{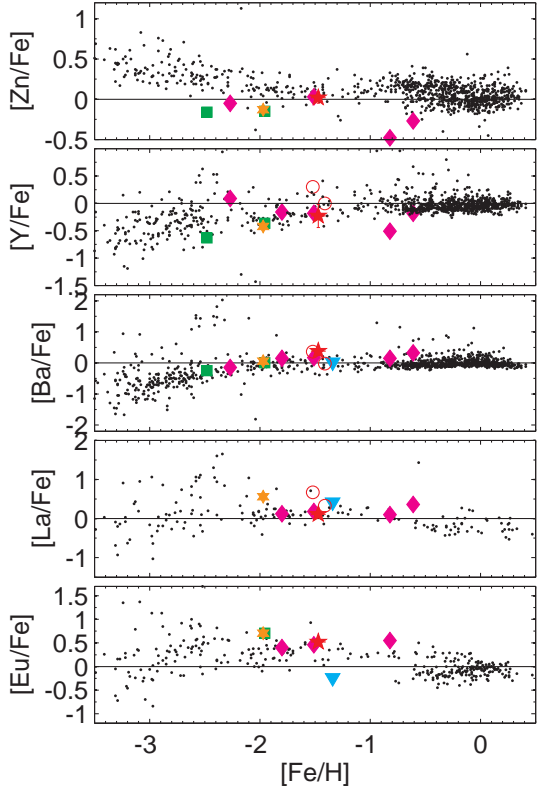}
\end{center}
\caption{Same as Fig.~3, but for neutron-capture elements. Eu disk abundances are from Koch \& Edvardsson (2002).}
\end{figure}
All of these elements have sound measurements, while [Eu/Fe] is based on a marginal detection (EW=17 m\AA) of the 6645~\AA~line. 
As for the remainder of the elements, also the neutron-capture elements we determined (Fig.~5)
are fully compatible with Galactic halo stars and GCs at similar metallicities. This indicates that Pal~5's early evolution was governed 
by the same enrichment processes as other, old and metal-poor GCs over a broad range of masses so that the 
subsequent strong mass-loss had no bearing on the chemical abundance patterns observed today. 
\subsection{ RGB versus AGB subsamples}
It has been  known that AGB stars can have lower metal abundances when using photometric
gravities in the abundance analysis, a fact that is commonly  attributed to the impact of departures from the LTE assumption 
 (Lapenna et al. 2015, and references therein). 
Thus we divided our samples into an AGB-only and a RGB-only sample and proceeded with EW measurements and abundance analyses in full analogy
to the methods outlined above (see also Koch \& C\^ot\'e 2010). These separate EW measurements and resulting abundance results are detailed in Tables~3 and 4. 
We note, however, the low SNR of the AGB sample and the values listed here should be used with caution.

The main effect is a lowering of the iron abundances as was already seen in the global metallicities (Sect.~5.1); upon this subdivision, however, an excellent ionization equilibrium for the AGB sample 
could be reached, which supports the suggestion that NLTE effects could be the driver of the abundance departures. 
 A pronounced difference is seen in the [Na/Fe] abundance ratios between AGB and RGB, with red giants having a markedly 
higher Na abundance by 0.5 dex. This may be due to different extents of the NLTE corrections in these types of stars.
{ Furthermore, this is just what is expected if a Na-rich second generation of stars (which are also more He-rich than 
first generation stars) fails to reach or complete the AGB phase. We discuss this further in the context of light element variations
within this GC in Sect.~7.} 
The opposite is true for Mg: here, a low Mg in the RGB sample is contrasted by the AGBs' strong enhancement (Sect. 7; see also Mucciarelli et al. 2012). 
The remainder of the measured elements agree to with the error bars, which are inevitably larger due to the lower SNR of the subsamples. 
\section{Comparison with other clusters}
The distinguishing feature of Pal~5 is clearly its severe state of tidal disruption. Nevertheless, all chemical abundance information
had already been imprinted on its stars at an early stage before this dynamical alteration began.
Thus, chemically speaking, it is a typical representative of the old, metal-poor halo population. This is highlighted in Figs.~3--5 through
a comparison to a selection of other GCs that are interesting for a variety of reasons. 

M5 is an inner halo (R$_{\rm GC}$=6 kpc) GC with a metallicity similar to that of Pal~5. 
It is presently much more massive than the very low-mass Pal~5 (at M$_V=-8.81$ vs. $-5.17$ mag) and shows only mild evidence of 
tidal distortion (Jordi \& Grebel 2010). While Smith et al. (2002) note a lower [$\alpha$/Fe] ratio in Pal~5 compared to M5, all other elements measured in that study and by us are fully compatible with those in M5.

In Figs.~3--5 we overplotted several objects that have been associated with Sagittarius. Given the complex star formation and enrichment history 
of this massive Galactic satellite, those GCs span a broad range of metallicities from $\sim -0.5$  to below $-2$ dex. 
In this regard, younger outer halo clusters such as Pal~12 ([Fe/]H=$-0.8$ dex; R$_{\rm GC}$=16 kpc) show the depleted [$\alpha$/Fe] ratios typical of environments with a low star formation efficiency. This value amounts to 0.06 dex for Pal~12 (Cohen 2014) and is 
significantly lower than found in Pal~5, despite  the latter already having an enhancement  lower than the halo average. 
Other such clusters, such as M~54 at the same metallicity as Pal~5, are very similar in most of the elements, again indicating that 
the dynamical history of these systems has not affected any of their basic chemical properties.  

Located beyond 90 kpc, Pal~3 and Pal~4 are amongst the most{ distant} GCs in the Milky Way halo. Despite their younger ages ($\sim$10 Gyr 
compared to 12 Gyr for other, typical old GCs; { Mar{\'{\i}}n-Franch et al. 2009; Dotter et al. 2011}), the main conclusion from the co-added abundance analyses of Koch et al. (2009) and Koch \& C\^ot\'e (2010) was that these outer halo clusters are  chemically indistinguishable from the older ones in the inner halo and have thus experienced 
similar enrichment histories.  Pal~3 and Pal 4 have metallicities comparable to that of Pal~5 and most of their chemical abundances
are very similar. In particular, Pal~3 has a lower mean [Mg/Fe] ratio that is  identical to the one we found for Pal 5. 
In a few cases, other element ratios deviate from the halo distribution and from Pal~5's abundances, such as 
 higher Sc and enhanced Co for both clusters, and elevated Zn and La abundances in Pal 3;  for the remaining elements, 
 all the Pal's (3, 4, and 5) element ratios agree within the uncertainties.  
Finally, we note that both of the studies mentioned above used the same technique of  co-adding spectra and 
yielded very similar abundance results and precisions.  

NGC 5466 is metal-poor ([Fe/H=$-1.97$ dex) and shows tidal tails extending $\sim4\degr$ (Belokurov et al. 2006).
An optical and infrared abundance analysis by Lamb et al. (2015) revealed typical GC abundance patterns for all 
studied elements in that this cluster shows light element variations and neutron-capture abundances indicative of pollution from AGB stars.  
Despite its state of disturbance, and a possible connection to the Sgr dwarf galaxy, Lamb et al. (2015) could not detect any 
obvious differences from the prevailing halo distribution at those low metallicities. NGC 5466's $\alpha$-elements are 
depleted to a similar extent as found for Pal~5, albeit at half a dex lower metallicity. 
We note that those authors used a combination of optical and infrared spectroscopy and accounted for departures from LTE 
for several elements, so that  slight differences to the literature halo and GC samples (performed in LTE from optical data) 
occur naturally. 

Another interesting reference object is the bulge cluster NGC~6712 at [Fe/H]=$-1.01$ dex (Yong et al. 2008). This 
object has lost a considerable amount of its mass as manifested in its mass and luminosity functions (de Marchi et al. 1999).
It also shows O and Na abundance variations as strong as in any other GC, but which are not seen 
in halo field stars (e.g., Geisler et al. 2007). From the sheer large extent of the O and Na abundance variations in their sample of 
five stars (with full range of 0.6 dex and 1$\sigma$ scatter of 0.28 dex both in Na and O), Yong et al. (2008) concluded that 
this GC must have been much more massive in the past in order to allow for these element variations to develop across
the cluster's multiple populations. Such large spreads were also found by Smith et al. (2002) in Pal~5,  which also has lost up to 90\% 
of its initial mass. Therefore, it is timely to ask whether such light element variations could be detected in an abundance analysis
based on co-added spectra. 
\section{Light element variations}
The wavelength range of our spectra allowed us, in principle, to estimate the carbon-richness of the stars based on 
the CH G-band at $\sim$4300 \AA. The derived CH-index 
 provides a well-calibrated tool measured on low-resolution spectra, which  often results in bimodal distributions 
indicative of the complex chemical enrichment processes in GCs  (e.g., Smith 1985; Kayser et al. 2008). 
While we could attempt to measure a similar band-index, we note that our spectra cut-off redwards of the bluest 
end of the CH-bandpass commonly used in such population studies (i.e., at 4298  \AA~instead of the  limit of 4285 \AA; 
e.g., Eq.~2 in Kayser et al. 2008). This renders the zeropoint of our measurements incommensurable.
It is, however, still very useful in identifying possible abundance spreads or outliers such as CH-strong or -weak stars. 

\begin{figure}[htb]
\begin{center}
\includegraphics[angle=0,width=1\hsize]{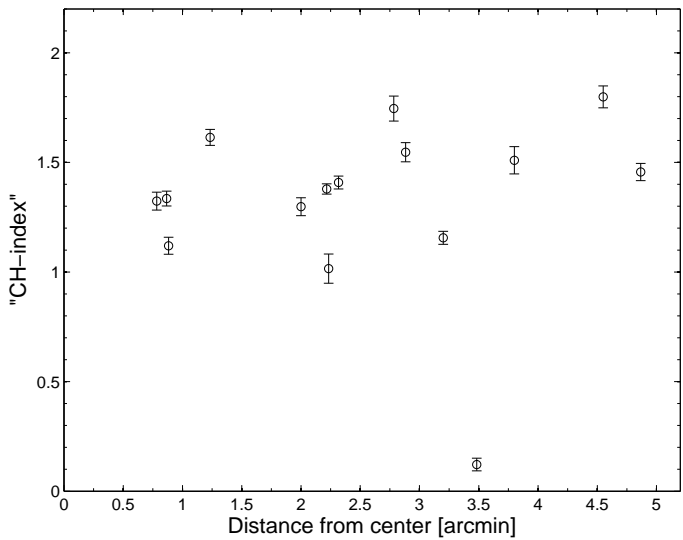}
\end{center}
\caption{Pseudo-CH-index versus position in the GC. The CH-weak AGB-star Pal-J clearly stands out.}
\end{figure}
As Fig.~6 indicates, one object (the AGB-star J at 3.5$\arcmin$ or 1.3 half-light radii) has a significantly lower 
CH-index indicating that it is carbon-weak. 
The CH band strength is usually anti-correlated with the N-abundance, which  is commonly quantified by CN-band index, in the sense that CH-weak stars tend to be CN-strong (e.g., Kayser et al. 2008; Martell et al. 2010), so we should expect star J to also have a significantly stronger 
CN band. However, as already noted by Smith (1985), this is not the case and star J is also CN-weak, which prompted him 
to invoke metallicity effects to explain the extraordinarily low C-abundance. { Indeed, star J is the member with the lowest metallicity in our sample.}

Our method explicitly assumes that none of the chemical elements shows any abundance spread.
Given the evidence of light-element (O, Na, Mg, Al) variations in every GC studied to date, this is clearly a false premise.
In order to test how our co-added abundance derivation responds to the presence of an abundance spread, 
 we emulated a Na-O anti-correlation by randomly populating this abundance space with 10$^5$ stars, following the dilution model for M 5 from Carretta et al. (2009). This GC  is very similar to Pal 5 in most respects (Ivans et al. 2001; Carretta et al. 2009). 
The ratio of first generation (low-Na, high-O) to second generation stars (high-Na, low-O)
was chosen as  30:70 (e.g., Carretta 2013). Out of this random sample, we  drew 15 random values (corresponding to the number of targets) and predicted individual EWs for each with MOOG, ultimately co-adding them as before (Sect.~4.2). 

As a result, we found no correlation  of the combined EW with the sampling of Na-poor and Na-rich stars from the mock cluster, as quantified by the  [Na/Fe]  interquartile range of the input sample.  
This means  that co-added EW measurements or syntheses of integrated-light spectra 
(as also in, e.g.,  Koch \& C\^ot\'e 2010; Sakari et al. 2013; Sch\"onebeck et al. 2014) 
 are rather insensitive to any spread in chemical elements, as exemplified here by sodium; even more so, 
 since for Pal~5, a large Na-spread of 0.6 dex is already known from  high-resolution measurements of  
 four individual stars (Smith et al. 2002).
 
 We note a trend in the Na and Mg abundances when performing an analysis on the AGB- and RGB-only subsamples,
 in that a  systematically lower Na of the AGB sample goes along with a high Mg-abundance (Sect.~5.6). 
 Conversely, the RGB-sample appears depleted in Mg and enhanced in Na. 
For want of a measurement of oxygen in either spectrum, we can take Mg as proxy for O as they are both produced via hydrostatic burning, albeit in 
different cycles that would render Mg-Al a more suitable comparison. In this simplistic scenario, 
our measured Mg and Na abundances would reflect the presence of light element variations in analogy to the Na-O anti-correlation, 
 if we were to presume that the AGB spectrum was dominated by first generation (high Mg, low Na) stars.
{ This is also in line with findings in some GCs that Na- and He-rich  second generation stars can fail to reach or complete their AGB-phase (e.g., McLean et al. 2016).}
\section{Summary and conclusions}
We determined chemical abundance ratios for various tracers of chemical evolution in the tidally disrupted GC Pal~5 using the co-addition 
technique we developed in earlier studies of remote Milky Way satellites. While this technique only allows us to measure the integrated properties of the stellar system, our results are fully compatible with results from a low number of individual  high-resolution spectra (four 
stars of Smith et al. 2002). As a result, we found  that Pal~5 is not unusual in any regard and that it follows the abundance trends of metal-poor GCs very closely, indicating that tidal perturbations over the course of Gyrs of evolution have no impact whatsoever on the 
chemical properties of these systems, { although the present-day mass of GCs remains one of the main drivers of the extent of the observed light element variations (Carretta et al. 2010)}.  

We were able to measure mean abundance ratios for 20 elements to high precision, however, 
our statistical tests have shown that the method of co-adding spectra is not sensitive to disentangling abundance variations such as the Na-O
anti-correlation, even if spreads of the  typical, high degree of 0.6 dex are present. To better characterize those, individual spectra of a large number of stars are still the most viable course. 
\begin{acknowledgements}
The authors thank Kyle Cudworth and Andrea Schweitzer for making their membership catalogue available to us. We are grateful to the  anonymous referee for a very fast and careful report.
This work was based on observations obtained at the W. M. Keck Observatory, which is operated jointly by the California Institute of Technology and the University of California. We are grateful to the W. M. Keck Foundation for their vision and generosity. We recognize the great importance of Mauna Kea to both the native Hawaiian and astronomical communities, and we are grateful for the opportunity to observe from this special place.
 \end{acknowledgements}
\end{document}